\documentclass{elsart}
\usepackage{epsfig}
\usepackage{amssymb}
\usepackage{amsmath}

\newcommand{\affuni}[2]{Dipartimento di Fisica dell'Universit\`a #1, #2, Italy.}
\newcommand{\affinfn}[2]{INFN Sezione di #1, #2, Italy.}
\newcommand{\dafne}     {DA$\Phi$NE }

\newcommand{\pio}{\mbox{$\pi^{0}$}}

\newcommand{\ks}{\mbox{$K_{S}$}}
\newcommand{\kl}{\mbox{$K_{L}$}}

\begin{document}

\begin{frontmatter}

\title{\boldmath  A new limit on the CP violating
  decay  $\ks \to 3 \pio$ with the KLOE experiment}

\collab{The KLOE-2 Collaboration}
\author[Frascati]{D.~Babusci},
\author[Roma2,INFNRoma2]{D.~Badoni},
\author[Cracow]{I.~Balwierz-Pytko},
\author[Frascati]{G.~Bencivenni},
\author[Roma1,INFNRoma1]{C.~Bini},
\author[Frascati]{C.~Bloise},
\author[Frascati]{F.~Bossi},
\author[INFNRoma3]{P.~Branchini},
\author[Roma3,INFNRoma3]{A.~Budano},
\author[Uppsala]{L.~Caldeira~Balkest\aa hl},
\author[Frascati]{G.~Capon},
\author[Roma3,INFNRoma3]{F.~Ceradini},
\author[Frascati]{P.~Ciambrone},
\author[Messina,INFNCatania]{F.~Curciarello},
\author[Cracow]{E.~Czerwi\'nski},
\author[Frascati]{E.~Dan\`e},
\author[Messina,INFNCatania]{V.~De~Leo},
\author[Frascati]{E.~De~Lucia},
\author[INFNBari]{G.~De~Robertis},
\author[Roma1,INFNRoma1]{A.~De~Santis},
\author[Roma1,INFNRoma1]{A.~Di~Domenico},
\author[Napoli,INFNNapoli]{C.~Di~Donato},
\author[INFNRoma2]{R.~Di~Salvo},
\author[Frascati]{D.~Domenici},
\author[Bari,INFNBari]{O.~Erriquez},
\author[Bari,INFNBari]{G.~Fanizzi},
\author[Roma2,INFNRoma2]{A.~Fantini},
\author[Frascati]{G.~Felici},
\author[Roma1,INFNRoma1]{S.~Fiore},
\author[Roma1,INFNRoma1]{P.~Franzini},
\author[Roma1,INFNRoma1]{P.~Gauzzi},
\author[Messina,INFNCatania]{G.~Giardina},
\author[Frascati]{S.~Giovannella},
\author[Roma2,INFNRoma2]{F.~Gonnella},
\author[INFNRoma3]{E.~Graziani},
\author[Frascati]{F.~Happacher},
\author[Uppsala]{L.~Heijkenskj\"old}
\author[Uppsala]{B.~H\"oistad},
\author[Frascati]{L.~Iafolla},
\author[Uppsala]{M.~Jacewicz},
\author[Uppsala]{T.~Johansson},
\author[Cracow]{K.~Kacprzak},
\author[Uppsala]{A.~Kupsc},
\author[Frascati,StonyBrook]{J.~Lee-Franzini},
\author[Frascati]{B.~Leverington},
\author[INFNBari]{F.~Loddo},
\author[Roma3,INFNRoma3]{S.~Loffredo},
\author[Messina,INFNCatania,CentroCatania]{G.~Mandaglio},
\author[Moscow]{M.~Martemianov},
\author[Frascati,Marconi]{M.~Martini},
\author[Roma2,INFNRoma2]{M.~Mascolo},
\author[Roma2,INFNRoma2]{R.~Messi},
\author[Frascati]{S.~Miscetti\corauthref{cor}},
\author[Frascati]{G.~Morello},
\author[INFNRoma2]{D.~Moricciani},
\author[Cracow]{P.~Moskal},
\author[INFNRoma3,LIP]{F.~Nguyen},
\author[INFNRoma3]{A.~Passeri},
\author[Energetica,Frascati]{V.~Patera},
\author[Roma3,INFNRoma3]{I.~Prado~Longhi},
\author[INFNBari]{A.~Ranieri},
\author[Mainz]{C.~F.~Redmer},
\author[Frascati]{P.~Santangelo},
\author[Frascati]{I.~Sarra},
\author[Calabria,INFNCalabria]{M.~Schioppa},
\author[Frascati]{B.~Sciascia},
\author[Cracow]{M.~Silarski\corauthref{cor}},
\author[Roma3,INFNRoma3]{C.~Taccini},
\author[INFNRoma3]{L.~Tortora},
\author[Frascati]{G.~Venanzoni},
\author[Warsaw]{W.~Wi\'slicki},
\author[Uppsala]{M.~Wolke},
\author[Cracow]{J.~Zdebik}
\corauth[cor]{Corresponding author}
\address[Bari]{\affuni{di Bari}{Bari}}
\address[INFNBari]{\affinfn{Bari}{Bari}}
\address[CentroCatania]{Centro Siciliano di Fisica Nucleare e Struttura della Materia, Catania, Italy.}
\address[INFNCatania]{\affinfn{Catania}{Catania}}
\address[Calabria]{\affuni{della Calabria}{Cosenza}}
\address[INFNCalabria]{INFN Gruppo collegato di Cosenza, Cosenza, Italy.}
\address[Cracow]{Institute of Physics, Jagiellonian University, Cracow, Poland.}
\address[Frascati]{Laboratori Nazionali di Frascati dell'INFN, Frascati, Italy.}
\address[Mainz]{Institut f\"ur Kernphysik, 
Johannes Gutenberg Universit\"at Mainz, Germany.}
\address[Messina]{Dipartimento di Fisica e Scienze della Terra dell'Universit\`a di Messina, Messina, Italy.}\address[Moscow]{Institute for Theoretical and Experimental Physics (ITEP), Moscow, Russia.}
\address[Napoli]{\affuni{``Federico II''}{Napoli}}
\address[INFNNapoli]{\affinfn{Napoli}{Napoli}}
\address[Energetica]{Dipartimento di Scienze di Base ed Applicate per l'Ingegneria dell'Universit\`a 
``Sapienza'', Roma, Italy.}
\address[Marconi]{Dipartimento di Scienze e Tecnologie applicate, Universit\`a ``Guglielmo Marconi", Roma, Italy.}
\address[Roma1]{\affuni{``Sapienza''}{Roma}}
\address[INFNRoma1]{\affinfn{Roma}{Roma}}
\address[Roma2]{\affuni{``Tor Vergata''}{Roma}}
\address[INFNRoma2]{\affinfn{Roma Tor Vergata}{Roma}}
\address[Roma3]{Dipartimento di Matematica e Fisica dell'Universit\`a 
``Roma Tre'', Roma, Italy.}
\address[INFNRoma3]{\affinfn{Roma Tre}{Roma}}
\address[StonyBrook]{Physics Department, State University of New 
York at Stony Brook, USA.}
\address[Uppsala]{Department of Physics and Astronomy, Uppsala University, Uppsala, Sweden.}
\address[Warsaw]{National Centre for Nuclear Research, Warsaw, Poland.}
\address[LIP]{Present Address: Laborat\'orio de Instrumenta\c{c}\~{a}o e F\'isica Experimental de Part\'iculas,
Lisbon, Portugal.}

\begin{abstract}

We have carried out a  new direct search for the CP violating decay $\ks$
$\to 3 \pio$ with 1.7 fb$^{-1}$ of e$^+$e$^-$ collisions collected by the KLOE 
detector at the $\Phi$-factory DA$\Phi$NE. We have searched for this
decay in a sample of  about 5.9$\times 10^{8}$ $\ks \kl$ events
tagging 
the $\ks$ by means of the $\kl$ interaction in the calorimeter 
and requiring six prompt  photons. 
With respect to our previous search, the  analysis 
has been improved  by increasing of a factor four the tagged sample
and by a more effective background rejection of fake $\ks$ tags and
spurious clusters.
We find no candidates in data and simulated background samples, while
we expect 0.12 standard model  events. Normalizing to the number of $\ks \to 2
\pio$ events in the same
sample, we set the upper limit on
$BR(\ks \to 3 \pio) \leq 2.6 \times 10^{-8}$ at 90\% C.L.,  five times lower than
the previous limit. We also set the upper limit 
on the $\eta_{000}$ parameter, $|\eta_{000}| \leq
0.0088$ at 90\% C.L., improving by a factor two the latest direct measurement.
\end{abstract}

\end{frontmatter}

\section{Introduction}
\label{Sec:Intro}
The decay $\ks\to 3\pio$ violates CP invariance and its observation would be the first
example of CP violation in $\ks$ decays. 
The parameter $\eta_{000}$,  the  ratio of 
\ks\ to \kl\ decay amplitudes, is defined as: 
$\eta_{000} = A(\ks \to 3\pio)/A(\kl \to 3 \pio)=
\epsilon + \epsilon'_{000}$, where $\epsilon$ indicates
the $\ks$ CP impurity and $\epsilon'_{000}$ 
the contribution of a direct CP-violating term. 
Since we expect $\epsilon'_{000}\;\ll\;\epsilon$ 
\cite{dambrosio}, it follows that $\eta_{000} \sim \epsilon$.  
In the Standard Model, therefore, 
BR($\ks \to 3 \pio ) \sim  1.9 \times 10^{-9}$,  to a relative accuracy better than 1\% .
The observation 
of such decay remains quite a challenge.

Previous searches follow two alternative methods: via a fit to the interference
pattern or via a direct search. The NA48 collaboration~\cite{NA48} has 
fit the  $\ks/\kl\to 3\pi^0$ interference pattern at small decay 
times  finding 
$\Re\,(\eta_{000})=-0.002\;\pm\;0.011_{\rm stat}\;\pm\;0.015_{\rm sys}$ 
and $\Im\,(\eta_{000})=-0.003\;\pm\;0.013_{\rm stat}\;\pm\;0.017_{\rm sys}$, 
corresponding to a limit on BR($\ks\to 3\pi^0) \leq 7.4 \times 10^{-7}$ at 90\% C.L. 
The best upper limit on $BR(\ks\to 3\pio)$ comes from the direct search 
performed by the KLOE experiment~\cite{kloe2008} based on
450~pb$^{-1}$ of collision data collected during 2001-2002. 
KLOE observed 2 candidates, and quoted a  limit on $BR(\ks\to 3\pio) \leq 1.2 \times 10^{-7}$ at 90\%
C.L.~\cite{KLOE_old}.
In this Letter, we present a twofold improvement of this search based
on a four times larger, and independent, data sample collected
in 2004-2005 and on  improved techniques used for background rejection.
\section{The KLOE detector}
The KLOE experiment operated from 2000 to 2006 at \dafne, the Frascati 
$\phi$-factory. DA$\Phi$NE~\cite{dafne} is an $e^+e^-$ collider running at a 
center-of-mass energy of $\sim 1020$~MeV, the mass of the $\phi$ meson. 
Equal energy positron and electron beams collide at an angle of $\pi$-25~mrad,
producing $\phi$ mesons nearly at rest.
The detector consists of a large cylindrical Drift Chamber (DC)~\cite{DCH},
surrounded by a lead scintillating fiber Electromagnetic Calorimeter
(EMC)~\cite{EMC} both immersed in an axial 0.52~T magnetic field produced by a
superconducting coil around the EMC. At the beams interaction point, IP,
the spherical beam pipe of 10~cm radius is made of a Beryllium-Aluminum alloy
of 0.5 mm thickness.
Low beta quadrupoles are located inside the detector at a distance of 
about $\pm$~50~cm  from the interaction region.
The drift chamber, 4~m in diameter and 3.3~m long, has 12582
all stereo drift cells with tungsten sense wires and is a really light
structure with an average thickness less than  $0.1 $ X$_0$,	
having the chamber shell made of carbon fiber-epoxy composite
with an internal wall of $\sim 1$ mm thickness, and filled with
a  gas mixture of 90\% helium, 10\% isobutane, 
to minimize $K_S$ regeneration and photon conversion. 
The spatial resolutions are $\sigma_{xy} \sim 150\ \mu$m and 
$\sigma_z \sim$~2 mm.
The momentum resolution is $\sigma(p_{\perp})/p_{\perp}\approx 0.4\%$.
The calorimeter covers 98\% of the solid angle and is
composed by a barrel  and two endcaps, for a total of 88 modules.
Each module is read out at both ends by photomultipliers for a total
of 2440 cells arranged in five layers. 
The energy deposits are obtained from the signal amplitude, while the
arrival times and particles impact points are obtained from the spatial
coordinates of the fired cell and the time differences. 
Cells close in time and space are grouped into energy clusters. 
The cluster energy $E$ is calculated as the sum of the cell energies,
while the cluster time $T$ and position $\vec{R}$ are energy weighted averages. 
Energy and time resolutions are parametrized as $\sigma_E/E = 5.7\%/\sqrt{E\ {\rm(GeV)}}$ 
and  $\sigma_t = 57\ {\rm ps}/\sqrt{E\ {\rm(GeV)}} \oplus100\ {\rm ps}$, 
respectively.
The trigger \cite{TRG} uses both calorimeter and chamber information.
In this analysis events are selected with the calorimeter trigger,
requiring two energy deposits with $E>50$ MeV for the barrel and $E>150$
MeV for the endcaps. 
Data are then analyzed by an event classification filter \cite{NIMOffline},
which selects and streams various categories of events in different
output files.

In this Letter, we refer only to data collected during 2004-2005 for an integrated luminosity
$\mathcal{L} = 1.7~\mathrm{fb}^{-1}$ with the most stable running conditions and the
best peak luminosity. 
A total of 5.1 billion $\phi$ mesons were produced, yielding  1.7~$ \times~10^{9}$ \ks\kl\ pairs. 
Assuming BR($\ks \to 3 \pio$) $\sim  1.9 \times 10^{-9}$ about 3 signal 
events are expected to have been produced.
\section{\boldmath Event selection}
\label{Sec:DataSample}
At \dafne\, the mean decay length of $K_L$, $\lambda_L$, is equal to
$~\sim 340$ cm  and about
50\% of $K_L$'s reach the calorimeter before 
decaying. A very clean $K_S$ tag is provided by the $K_L$ interaction in the calorimeter 
($\kl$-crash), which is identified by a cluster with polar angle
$40^\circ<\theta_{cr}<140^\circ$, not associated
to any track, with energy $E_{cr}>100$~MeV and with 
a time corresponding to a $K_L$ velocity in the $\phi$ rest frame $\beta^*$
in the range [0.17,0.28].
The average value of the $e^+e^-$ center of mass 
energy $W$ is obtained with a precision of 20 keV  
for each 200 nb$^{-1}$  running period 
using large angle Bhabha scattering events~\cite{kloe2008}. The value of $W$ and 
the $K_L$-crash cluster position allows us to obtain, 
for each event, the direction of the $K_S$ with an 
angular resolution of 1$^{\circ}$ and a momentum 
resolution of about 2 MeV.

Because of its short decay length, $\lambda_S \sim 0.6$ cm, the displacement
of the $K_S$ from the $\phi$ decay position 
is negligible. We therefore identify as photons from $\ks$ decay,
neutral particles that travel with $\beta=1$ 
from the interaction point to the EMC (``prompt photons").
In order to retain a large control sample for the 
background while preserving high efficiency for the 
signal, we keep all photons satisfying $E_\gamma>$ 7 ~MeV and
$|\cos \theta|<$ 0.915.
Each cluster is required to satisfy the condition 
$|t_{\gamma}-R_{\gamma}/c|<{\rm min}(3.5\sigma_t, 2\ {\rm ns})$, 
where $t_{\gamma}$ is the photon flight time and $R$ the path 
length; $\sigma_t$ also includes a contribution from 
the finite bunch length (2--3 cm), which introduces 
a dispersion in the collision time. 
The photon detection efficiency of the calorimeter amounts to about 90\% for $E_\gamma$ = 20 MeV, and reaches 
100\% above 70 MeV. After tagging the signal sample is selected requiring 6 prompt photons.
For normalization we use the $K_S \to 2\pi^0$ decay which is selected requiring 4 prompt photons.\\
For both channels the expected background as well as the detector acceptance and the analysis efficiency
are estimated using the 
Monte Carlo simulation of the experiment~\cite{NIMOffline}.
The simulation incorporates a detailed geometry and material composition of the KLOE apparatus
and most of the data taking conditions of the experiment e.g. DA$\Phi$NE background rates,
position of the interaction point and beam parameters. All the processes contributing to
the background were simulated with statistics twice larger than
the data sample. Moreover, for the acceptance and the analysis efficiency evaluation a dedicated
$K_{S}\to 3\pi^{0}$ signal simulation was performed,  based on a branching
ratio equal to the best known upper limit~\cite{KLOE_old} increased by a factor of 30 (about 5000 events).
\subsection{The six-photon sample}
\begin{figure}
\begin{center}
\includegraphics[width=0.49\textwidth]{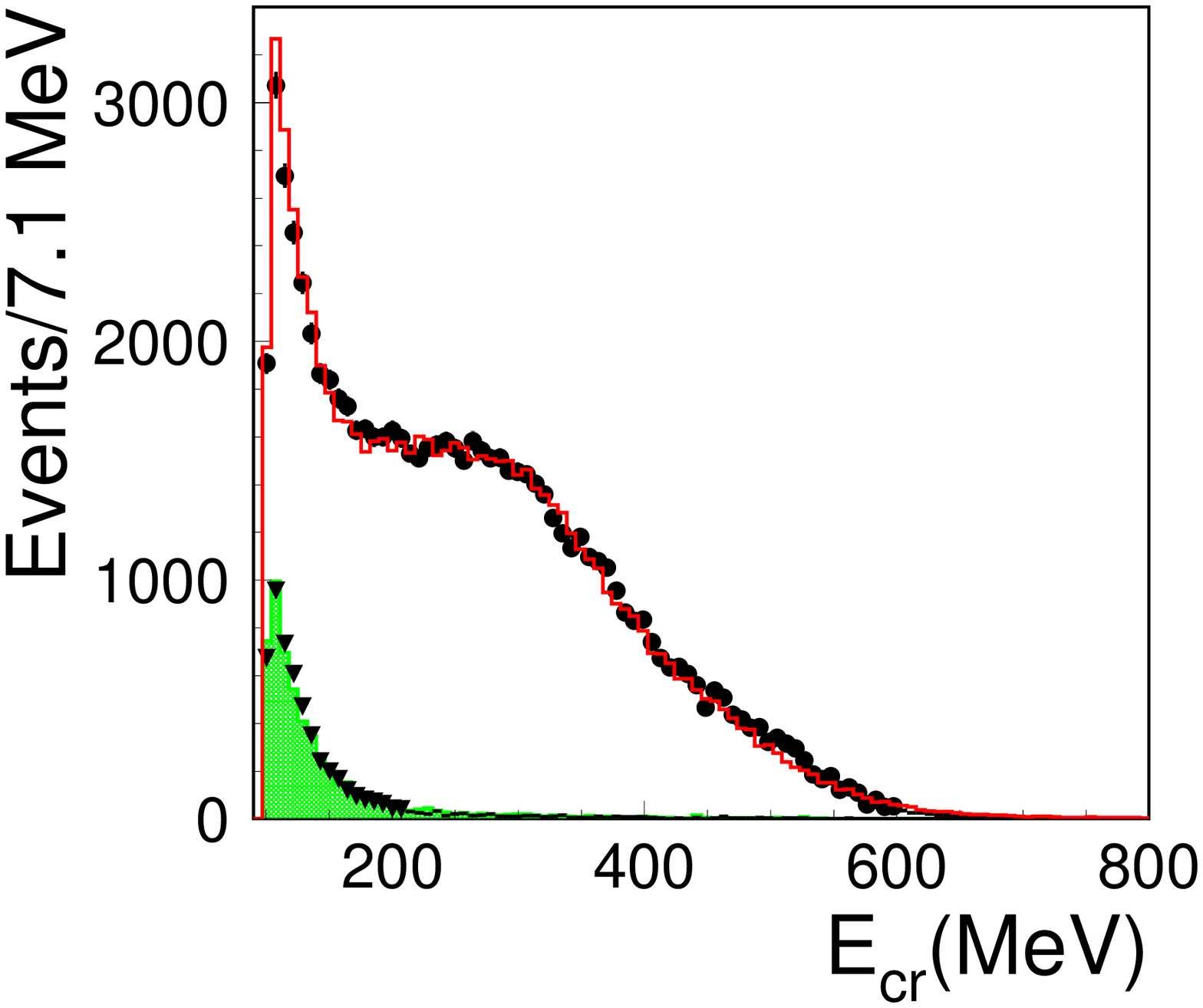}
\includegraphics[width=0.49\textwidth]{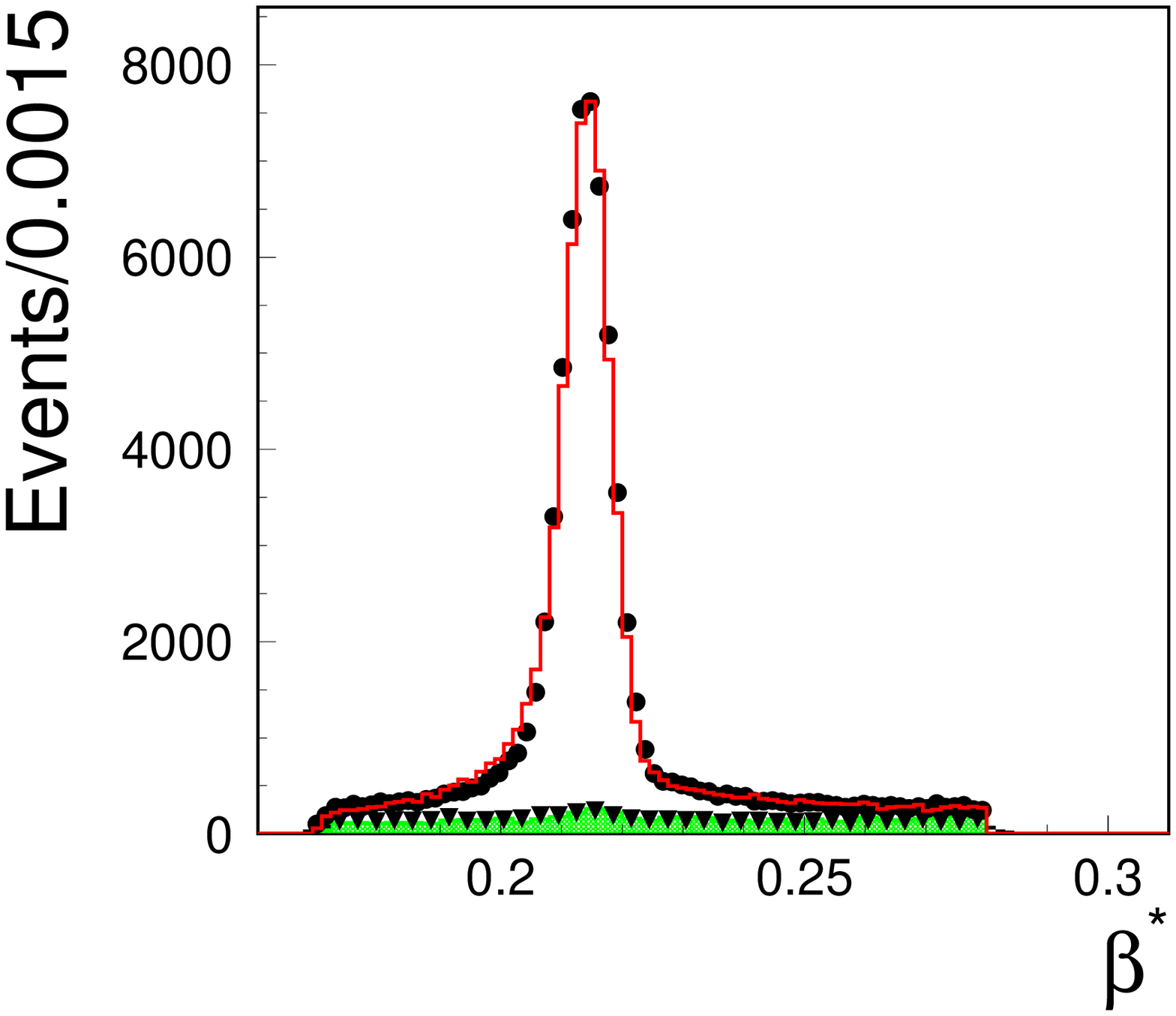}
\end{center}
\caption{Distributions of the $K_L$ energy deposit in the EMC ($E_{cr}$) and velocity
in the $\phi$ center of mass frame ($\beta^*$) for all events in the six-photon
sample. Black points represent data, while the MC background simulation is shown
as red histogram. The same distributions for events rejected
by the track veto are shown by the black triangles (data) and green filled histograms 
(MC simulation).}
\label{fig1}
\end{figure}
The selection of the $K_{S}\to 3\pi^{0}$ decay is performed by asking for a
$K_L$-crash and by searching six prompt photons from the decay of pions.
After these requirements we count 76689 events. For these events we perform
further discriminant analysis to increase the signal to background ratio.\\
The first analysis step aims to reject fake $K_S$ tags (about 2.5\% of the total background).
The distributions of $E_{cr}$ and $\beta^*$ for the selected data sample
and background simulations are shown in Fig.~\ref{fig1}. In the $\beta^*$ distribution,
the peak around 0.215 corresponds to genuine $K_L$ interaction in the calorimeter,
while the flat distribution mainly originates from
$\phi \to K_{S}K_{L} \to (K_S \to \pi^+\pi^-, K_L \to 3\pi^0)$ background events.
In this case one of the low momentum charged pions spirals in the forward direction and interacts
in the low-$\beta$ 
quadrupoles. This interaction produces neutral particles which simulate the signal of $K_L$
interaction in the calorimeter (fake $K_L$-crash), while the $K_L$ meson decays close
enough to the interaction point to produce six prompt photons.
To suppress fake $K_L$-crash we first reject events having charged particles produced close to
the interaction region (track veto).
The distributions of the kinematical variables for the vetoed
background events are shown in Fig.~\ref{fig1}.
Taking advantage of the differences in the $\beta^*$ and $E_{cr}$ distributions between
the tagged $K_S$ events and the fake $K_L$-crash, we have tightened
the cuts on these variables: $E_{cr} > 150~\mathrm{MeV}$ and $0.20 <\beta^* < 0.225$ ($K_L$-crash hard).
This improves by a factor 12 the rejection of this background with respect to the previous
analysis~\cite{KLOE_old}.\\
The second source of background originates from wrongly reconstructed $K_S\to 2\pi^0$ decays.
The four photons from this decay can be reconstructed as six due to fragmentation of the electromagnetic
showers (splitting). These events are characterized by one or two low-energy clusters reconstructed very close to
the position of the genuine photon interaction in the calorimeter and constitute about 67.5\% of the background.
Additional clusters come from accidental time coincidence between $\phi$ decay
and machine background photons from DA$\Phi$NE ($\sim$~30\% of the background).
\begin{figure}
\begin{center}
\includegraphics[width=0.49\textwidth]{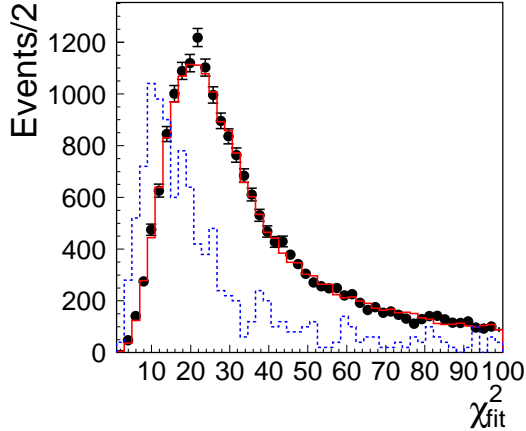}
\end{center}
\caption{
Distribution of $\chi^2_{fit}$ for the tagged sixphoton sample for data
(black points), background simulation (solid histogram), and simulated $K_S \to 3\pi^0$ signal
(dashed histogram).}
\label{fig:chi2}
\end{figure}
After tagging with the $K_L$-crash hard algorithm and applying the track veto we remain
with a  sample of about 50000 six-photon events.
A kinematic fit with 11 constraints has been performed imposing
energy and momentum conservation, the kaon mass and the velocity of the six photons in the final state.
The $\chi^2$ distribution of the fit for data and background simulation, $\chi^2_{fit}$, is shown in
Fig.~\ref{fig:chi2} together with the expected distribution for signal events.
Cutting on $\chi^2_{fit}$ reduces by about 30\% the remaining background while keeping the signal efficiency
at 70\% level.
\\
In order to improve rejection of events with split and accidental clusters, we have exploited
the correlation between two $\chi^2$-like variables named $\zeta_{2\pi}$ and $\zeta_{3\pi}$. 
$\zeta_{2\pi}$ is calculated by an algorithm selecting the best four out of six clusters satisfying
the kinematic constraints of the two-body decay in the $K_S \to 2\pi^0 \to 4\gamma$
hypothesis:
\begin{figure}
\begin{center}
\includegraphics[width=0.49\textwidth]{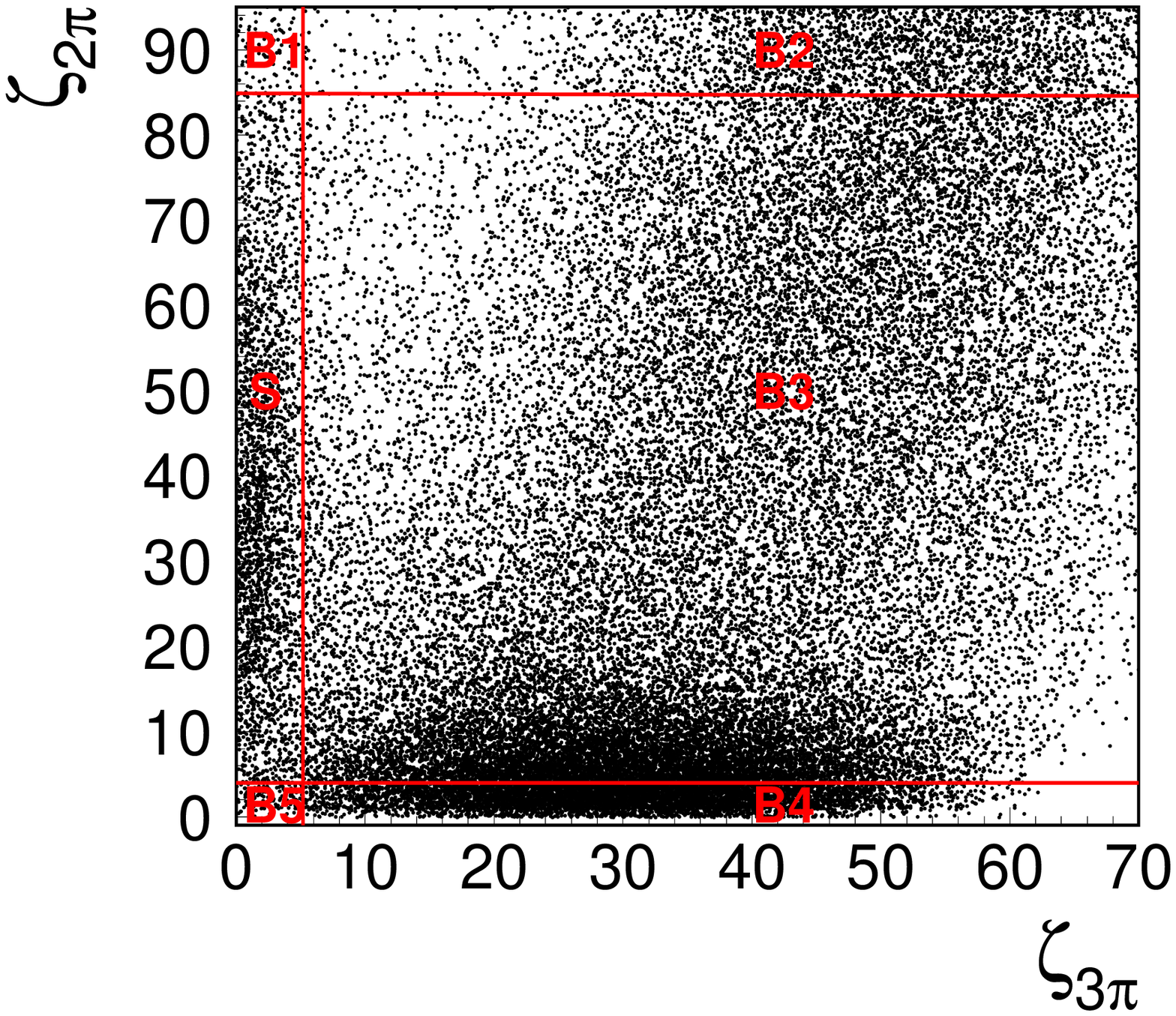}
\includegraphics[width=0.49\textwidth]{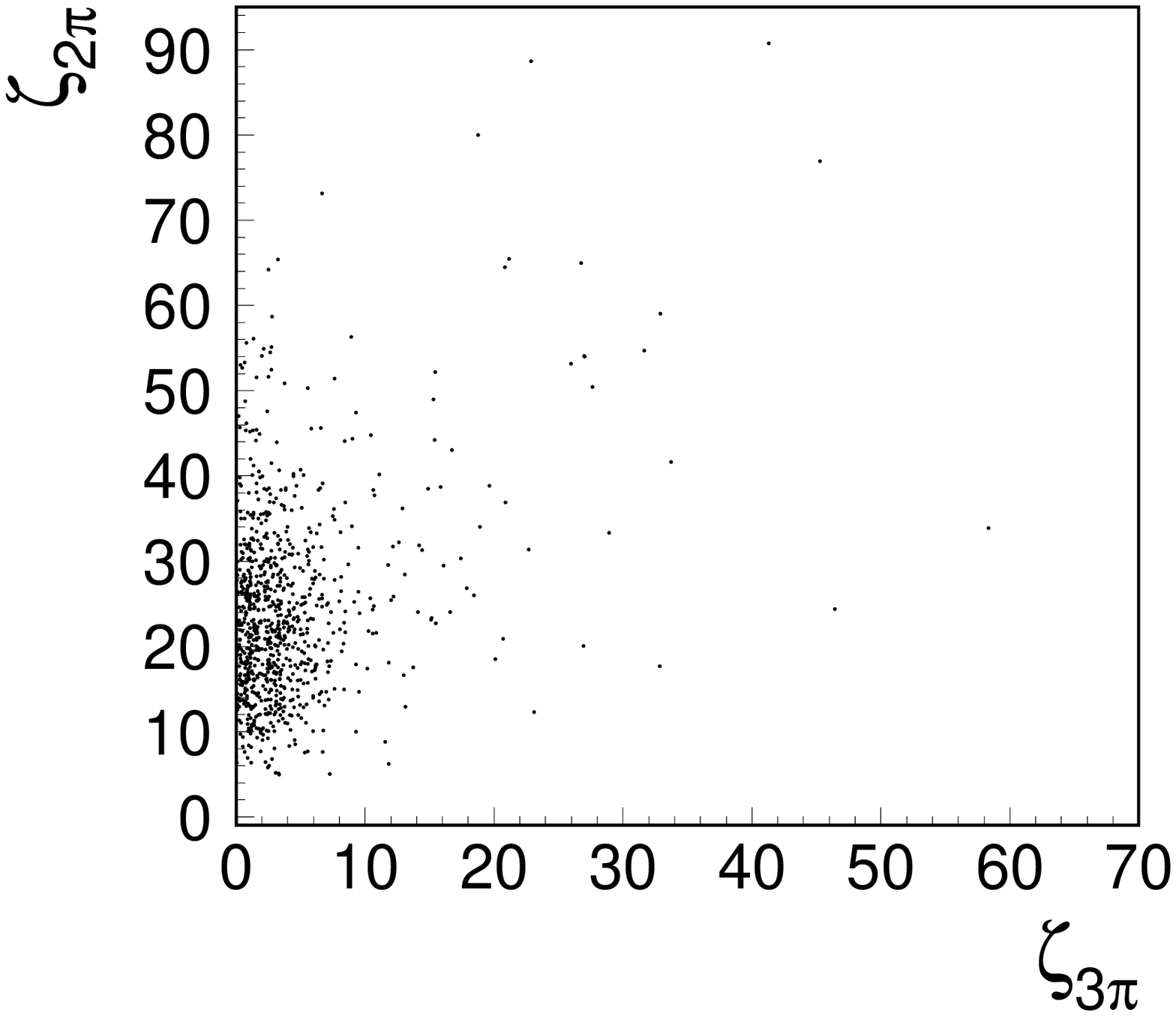}
\end{center}
\caption{Distributions of events in the $\zeta_{3\pi}$-$\zeta_{2\pi}$ plane,
for six-photon sample tagged by $K_L$-crash for data (left),
and for the simulated $K_S\to 3\pi^0$ decays (right).
The boundaries of the background control regions B1, B2, B3, B4, B5 and
the signal region S are as specified in the text.}
\label{fig3}
\end{figure}
\begin{align}
\zeta_{2\pi} &=~\frac{(m_{1 \gamma\gamma} - m_{\pi^0})^2}{\sigma^{2}_{2\pi}}
+ \frac{(m_{2 \gamma\gamma} - m_{\pi^0})^2}{\sigma^{2}_{2\pi}}
+ \frac{(\theta_{\pi\pi} - \pi)^2}{\sigma^2_{\theta_{\pi\pi}}}
\nonumber
+ \frac{\biggl(E_{K_{S}} - \displaystyle\sum_{i=1}^4 E_{\gamma_{i}}\biggr)^2}{\sigma^2_{E_{K_{S}}}}\\
&+ \frac{\biggl(p_{K_{S}}^x - \displaystyle\sum_{i=1}^4 p_{\gamma_{i}}^x\biggr)^2}{\sigma^2_{p_x}}
+ \frac{\biggl(p_{K_{S}}^y - \displaystyle\sum_{i=1}^4 p_{\gamma_{i}}^y\biggr)^2}{\sigma^2_{p_y}}
+ \frac{\biggl(p_{K_{S}}^z - \displaystyle\sum_{i=1}^4 p_{\gamma_{i}}^z\biggr)^2}{\sigma^2_{p_z}}~,
\label{chi2_2pi_def}
\end{align}
where $m_{1 \gamma \gamma}$ and $m_{2 \gamma \gamma}$ are the reconstructed
$\gamma \gamma$ masses for a given cluster pairing, and $\theta_{\pi\pi}$ denotes the opening angle
of the reconstructed pion directions in the $K_S$ center of mass frame. $E_{K_S}$ and $p_{K_S}$
stand for the $K_S$ energy and momentum vector determined from the reconstructed four-momentum
of $K_L$, while $E_{\gamma_i}$ and $p_{\gamma_i}$ are energies and momenta of four out
of six reconstructed photons.
The minimization of $\zeta_{2\pi}$ gives the best two photon pairs fulfilling the
$K_S \to 2\pi^0 \to 4\gamma$ hypothesis.
The resolutions used in Eq.~\ref{chi2_2pi_def} were estimated independently on data
and MC simulation using a $K_S \to 2\pi^0 \to 4\gamma$ control sample.\\
The second $\chi^2$-like variable, $\zeta_{3\pi}$, instead verifies the signal hypothesis $K_S \to 3\pi^0$ by looking
at the reconstructed masses of the three pions. For each pair of clusters
we evaluate $\zeta_{3\pi}$ as:
\begin{equation}
\zeta_{3\pi}~=~\frac{(m_{1 \gamma\gamma} - m_{\pi^0})^2}{\sigma^{2}_{3\pi}}
+ \frac{(m_{2 \gamma\gamma} - m_{\pi^0})^2}{\sigma^{2}_{3\pi}}
+ \frac{(m_{3 \gamma\gamma} - m_{\pi^0})^2}{\sigma^{2}_{3\pi}}~.
\label{chi2_3pi_def}
\end{equation}
As the best combination of cluster pairs, we take the configuration minimizing $\zeta_{3\pi}$.
The resolution on the $\gamma\gamma$ invariant mass in the $3\pi^0$ hypothesis, $\sigma_{3\pi}$,
was estimated applying the algorithm to the simulated $K_S \to 3\pi^0$ events.\\
The distributions in the $\zeta_{3\pi}$-$\zeta_{2\pi}$ plane for the data and
$K_S \to 3\pi^0$ simulated signal are shown in Fig.~\ref{fig3}. Signal events
are characterized by small values of $\zeta_{3\pi}$ and relatively high
$\zeta_{2\pi}$. To compare data and Monte Carlo simulations we have subdivided
the $\zeta_{3\pi}$-$\zeta_{2\pi}$
plane into six regions B1, B2, B3, B4, B5, and S as indicated in the left panel of Fig.\ref{fig3}.
Region S, with the largest signal-to-background ratio, is the signal box,
while B1--B5 are control regions used to check the reliability of the simulation
and optimize our description of the experimental data.\\
Simulation does not reproduce accurately the absolute number of events belonging
to different background categories. However, their kinematical properties
are reproduced quite well. To determine the background composition,
and improve the description of experimental data, we have performed a binned
likelihood fit of a linear combination 
of simulated $\zeta_{3\pi}$-$\zeta_{2\pi}$ distributions to the same data distribution
for all background categories.
The quality of the fit was controlled by comparing inclusive distributions
of discriminating variables between data and simulation. Examples are presented
in Fig.~\ref{fig4}.\\
\begin{table}
\begin{center}
\begin{tabular}{|c|c|c|c|c|c|c|}
\hline
\textbf{} & \textbf{SBOX} & \textbf{B1} & \textbf{B2} & \textbf{B3} & \textbf{B4} & \textbf{B5}\\
\hline
\textbf{DATA} & 220 $\pm$ 15 & 5 $\pm$ 3 & 15179 $\pm$ 123 & 26491 $\pm$ 163 & 6931 $\pm$ 83 & 137 $\pm$ 12\\
\hline
\textbf{MC} &  239 $\pm$ 11 & 4 $\pm$ 3 & 14905 $\pm$ 116 & 26964 $\pm$ 169 & 6797 $\pm$ 76 & 100 $\pm$ 7\\
\hline
\end{tabular}
\end{center}
\caption{
\label{tab:box1}
Number of events populating control regions in the $\zeta_{3\pi}$-$\zeta_{2\pi}$ plane defined in
Fig.~\ref{fig3} after tight requirements on $K_L$-crash and track veto.}
\end{table}
Table~\ref{tab:box1} shows the comparison of observed number of events with the expectations in each
control region of the $\zeta_{3\pi}$-$\zeta_{2\pi}$ plane. The agreement is better
than 1.5 $\sigma$ in all regions except region B5 (2.8 $\sigma$).\\
%
\begin{figure}
\begin{center}
\includegraphics[width=0.49\textwidth]{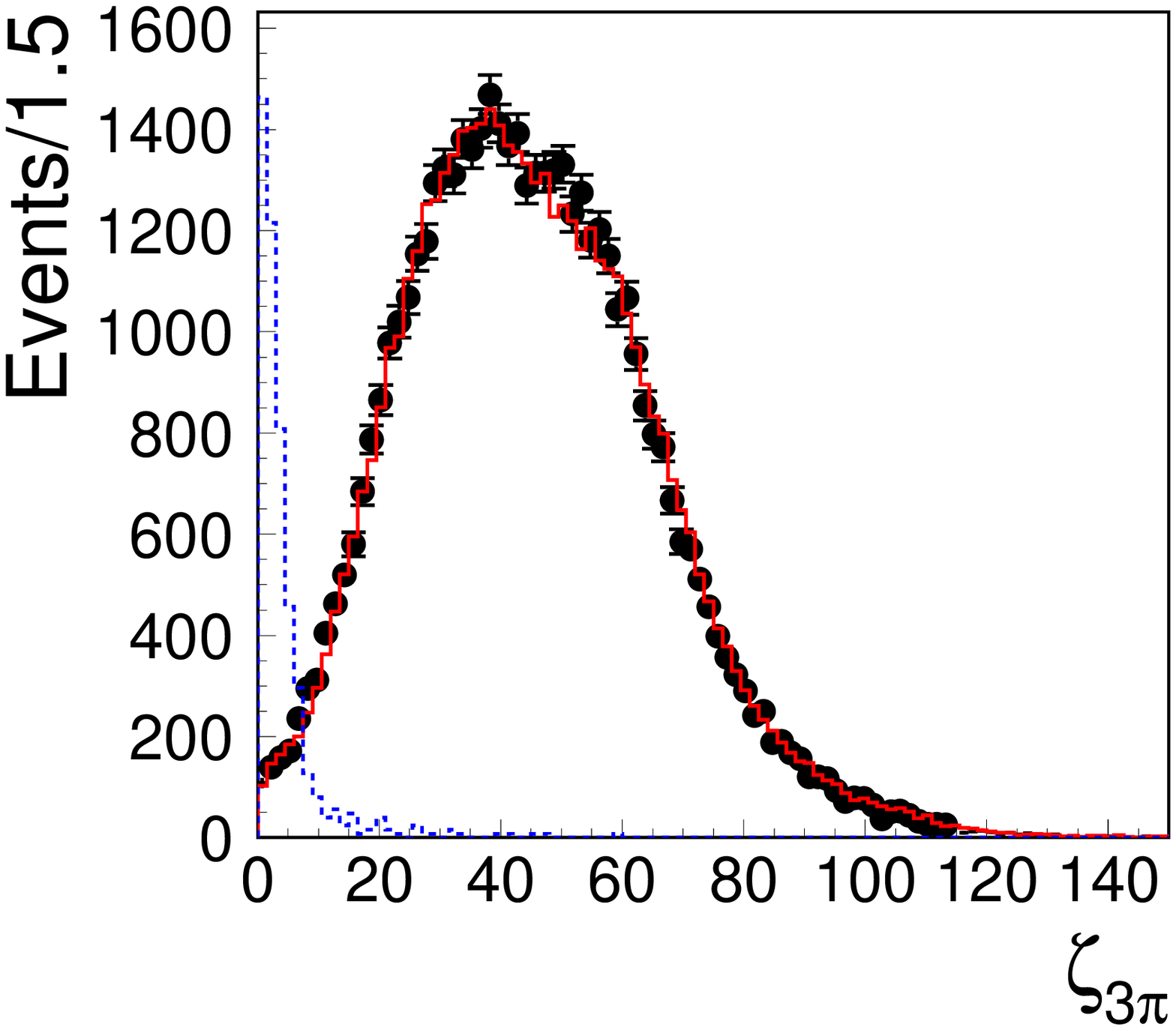}
\includegraphics[width=0.49\textwidth]{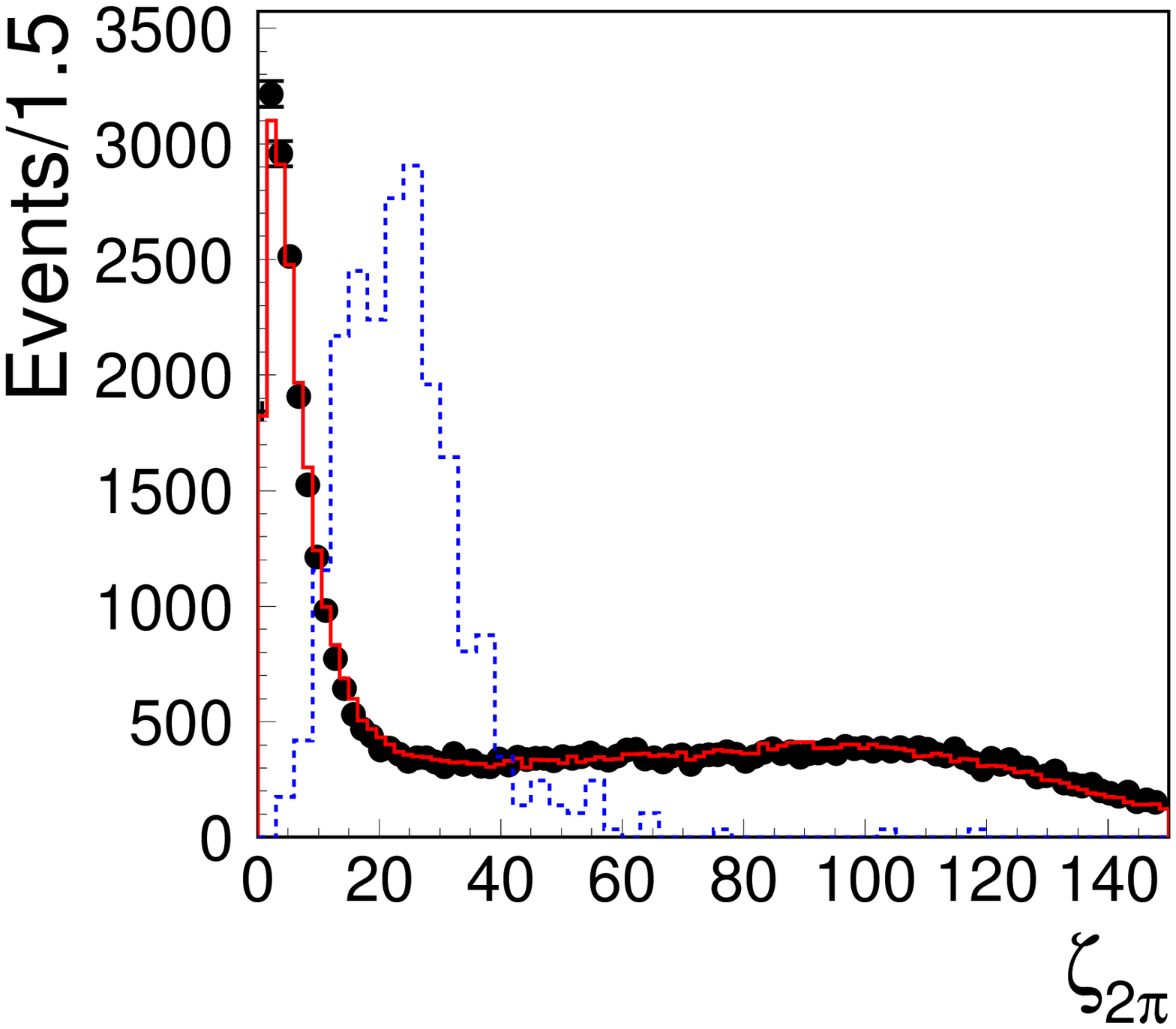}
\end{center}
\caption{Inclusive distributions of the $\zeta_{3\pi}$
and $\zeta_{2\pi}$ discriminating variables for six-photon events: data
(black points), background simulations (red curves). The dashed histograms
represents simulated $K_S \to 3\pi^0$ events.
}
\label{fig4}
\end{figure}
To further improve the $K_S \to 2\pi^0$ background rejection we cut on the $\Delta$ variable defined as:
\begin{equation}
\Delta~=~(m_{\phi}/2 - \sum E_{\gamma_{i}})/\sigma_{E}~,
\label{eqdE}
\end{equation}
where $\sum E_{\gamma_{i}}$ is the sum of energies of the four prompt photons selected by
the $\zeta_{2\pi}$ algorithm and $\sigma_E$ stands for the 4$\gamma$ energy resolution estimated using the
$K_S \to 2\pi^0 \to 4\gamma$ control sample. For $K_S \to 2\pi^0$ decays with two additional background
clusters, we expect $\Delta \sim$~0, while for $K_S \to 3\pi^0$ events $\Delta~\sim~m_{\pi^0}/\sigma_E$.
To further reject surviving $K_S \to 2\pi^0$ events with split clusters,
we cut on the minimal distance between centroids of
reconstructed clusters, $R_{min}$, considering that the distance between split clusters
is on average smaller than the distance between clusters originating from $\gamma$'s of
$K_S \to 3\pi^0$ decay.
Distributions of these two discriminant variables are presented
in Fig.~\ref{fig5}.\\
\begin{figure}
\begin{center}
\includegraphics[width=0.49\textwidth]{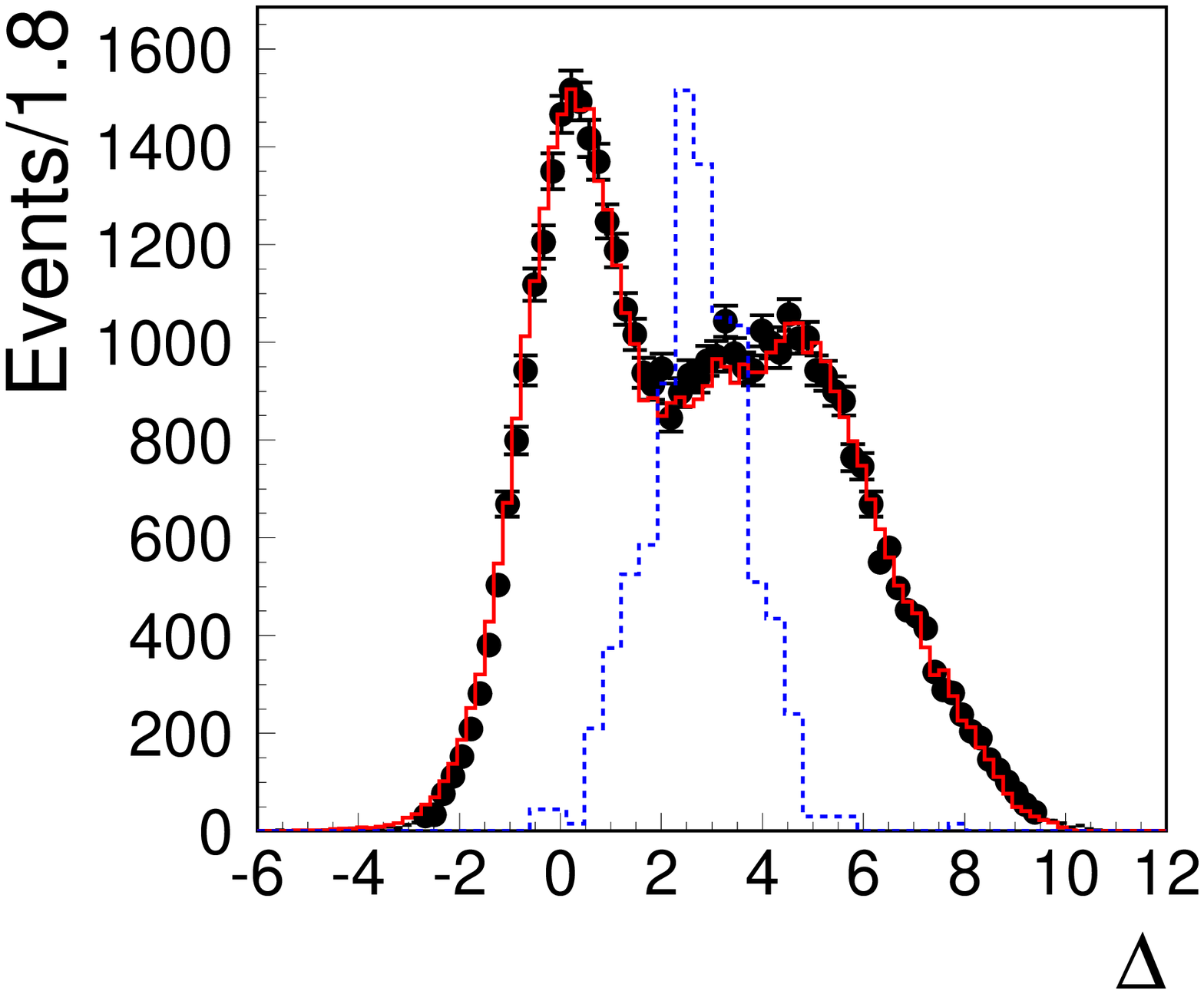}
\includegraphics[width=0.49\textwidth]{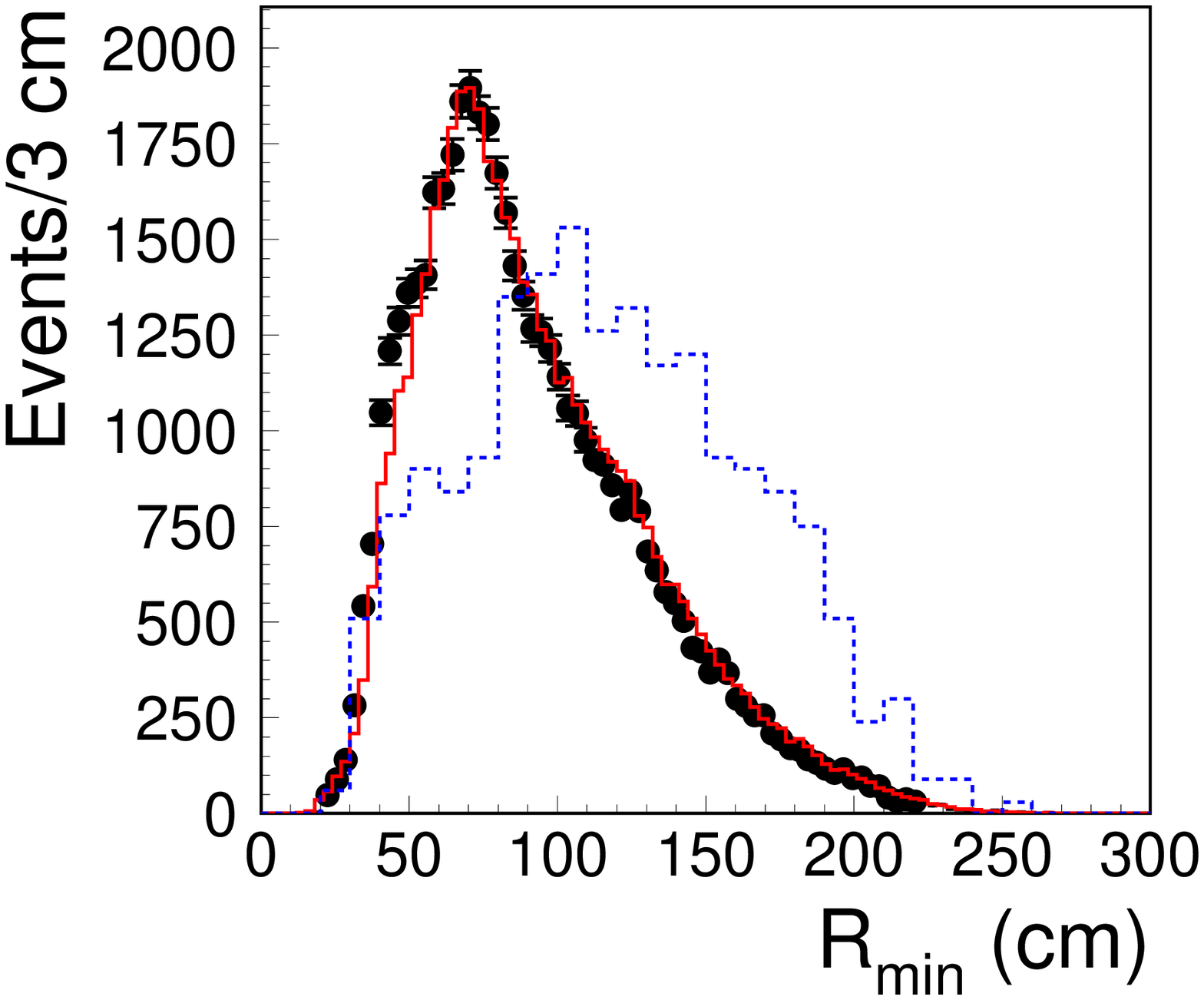}
\end{center}
\caption{Distributions of  $\Delta$
and $R_{min}$ discriminating variables for six-photon events:
data (black points), background simulations (red curves).
The dashed histograms represents simulated $K_S \to 3\pi^0$ events. 
}
\label{fig5}
\end{figure}
Before opening the signal box, the cuts on the discriminant variables have been refined minimizing
$f_{cut}(\chi^2_{fit},\zeta_{2\pi},\zeta_{3\pi},\Delta,R_{min}) = N_{up}/\epsilon_{3\pi}$, where
$\epsilon_{3\pi}$ stands for the signal efficiency and
$N_{up}$ is the mean upper limit (at 90\% CL) on the expected number of signal events calculated
on the basis of the expected number of background events
$B_{exp} = B_{exp}(\chi^2_{fit},\zeta_{2\pi},\zeta_{3\pi},\Delta,R_{min})$
from simulation~\cite{lal92}. The outcome of the optimizing procedure is $\chi^2_{fit} < 57.2$, 
$\Delta > 1.88$ and $R_{min} > 65$~cm. The signal box is defined as:
$4 < \zeta_{2\pi} < 84.9$ and $\zeta_{3\pi} < 5.2$.
At each stage of the analysis we checked that the simulation describes the data within
statistical uncertainty.
Distributions of $\chi^2_{fit}$, $\Delta$ and $R_{min}$ variables are presented
in Fig.~\ref{fig6} and Fig.~\ref{fig7} for events in the signal box.
In the right panel of Fig.~\ref{fig7} we present also the $R_{min}$ distribution just before
the last cut $R_{min}>65$~cm.
According to the Monte Carlo simulation, these survived events are all $K_S \to 2\pi^0$
decays with two split clusters (95$\%$), or one split and one accidental cluster (5$\%$).
A total efficiency of $\epsilon_{3\pi} = 0.233 \pm 0.012_{stat}$ has been estimated.  
\begin{figure}
\begin{center}
\includegraphics[width=0.49\textwidth]{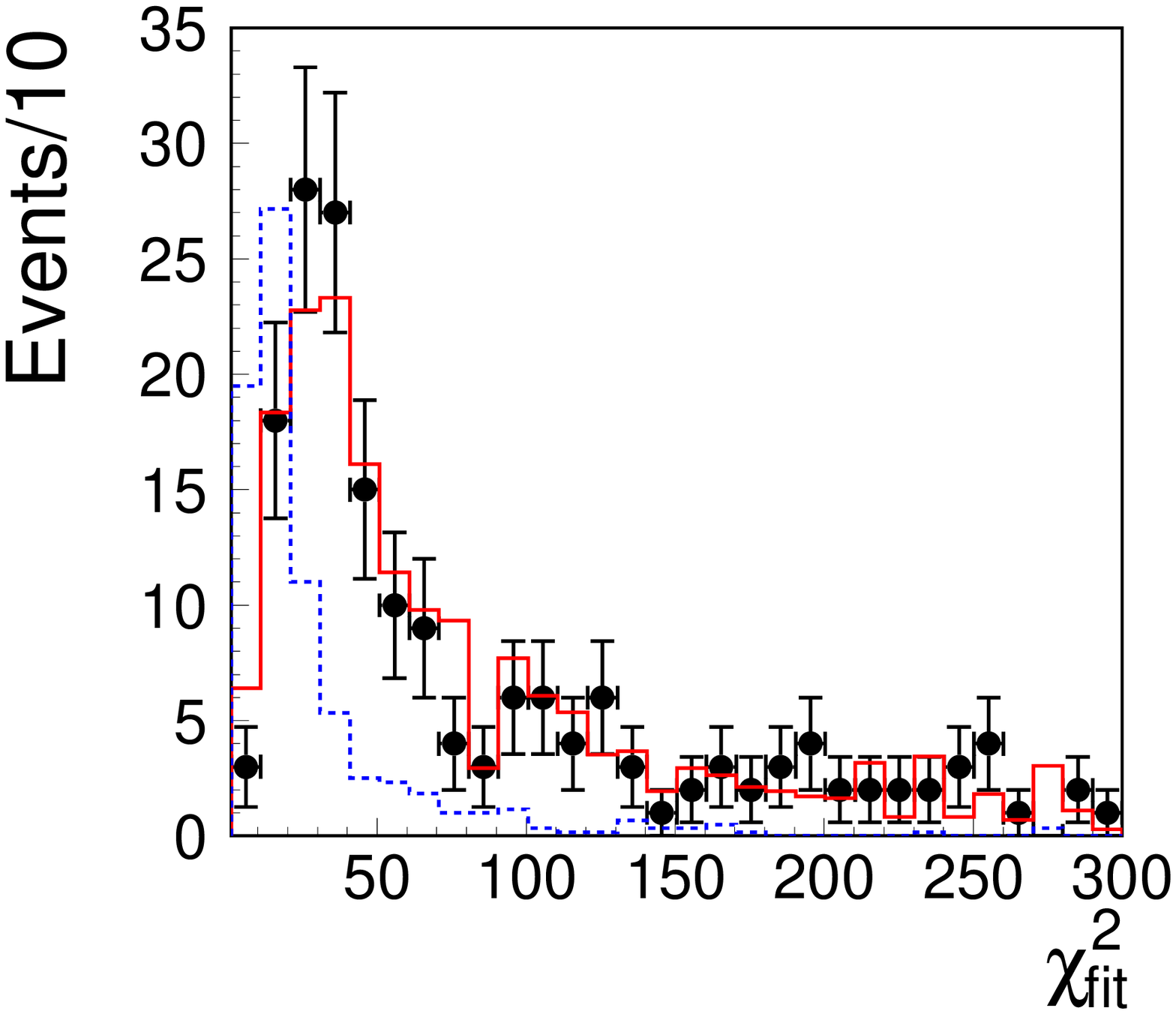}
\includegraphics[width=0.49\textwidth]{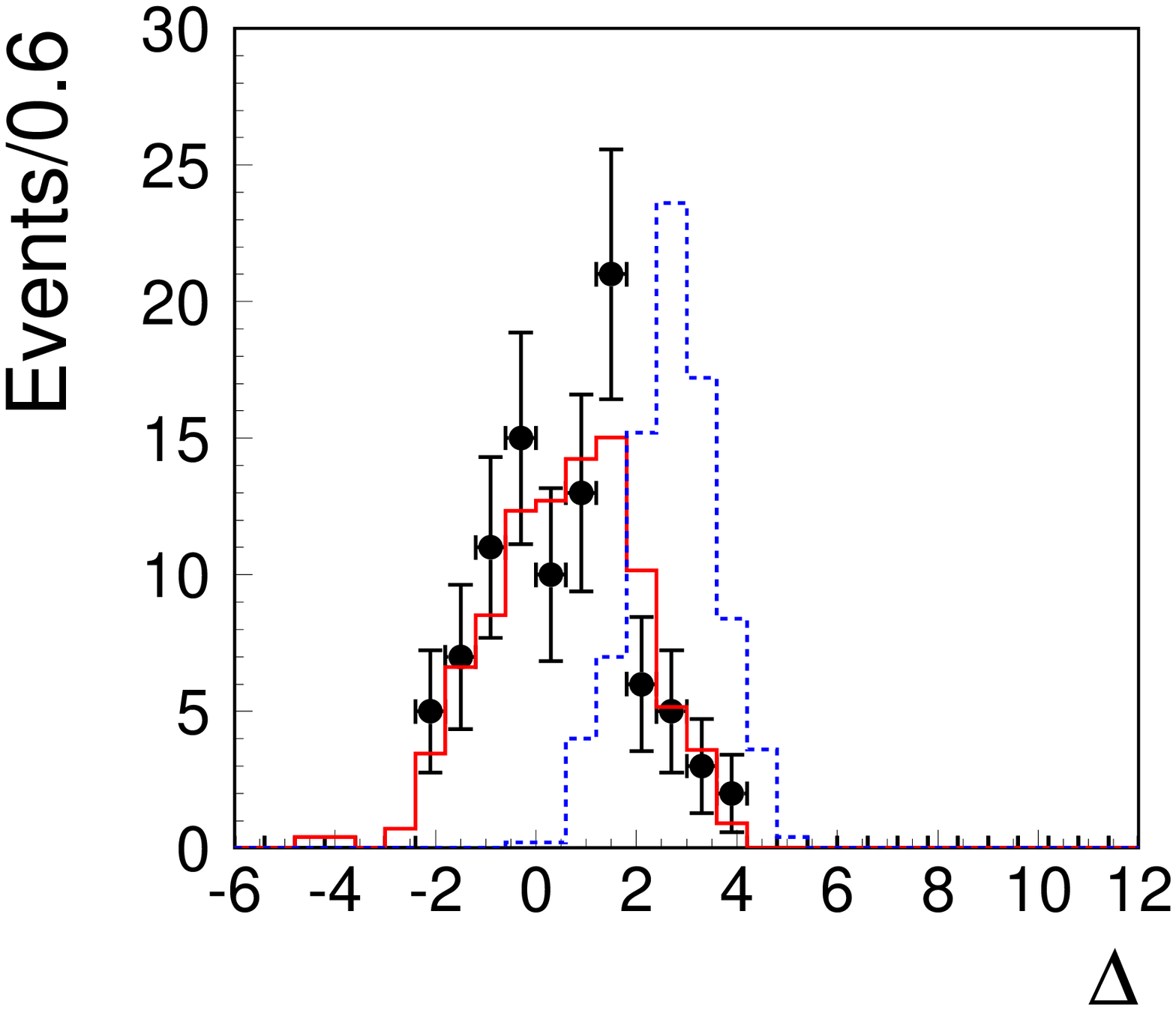}
\end{center}
\caption{Distributions of $\chi^2_{fit}$ for six-photon events in the signal box (left)
and $\Delta$ for six-photon events in the signal box applying the $\chi^2_{fit} <$~57.2 cut (right).
Black points are data, background simulation is the red histogram.
The dashed histogram represents simulated $K_S \to 3\pi^0$ events.}
\label{fig6}
\end{figure}
\begin{figure}
\begin{center}
\includegraphics[width=0.49\textwidth]{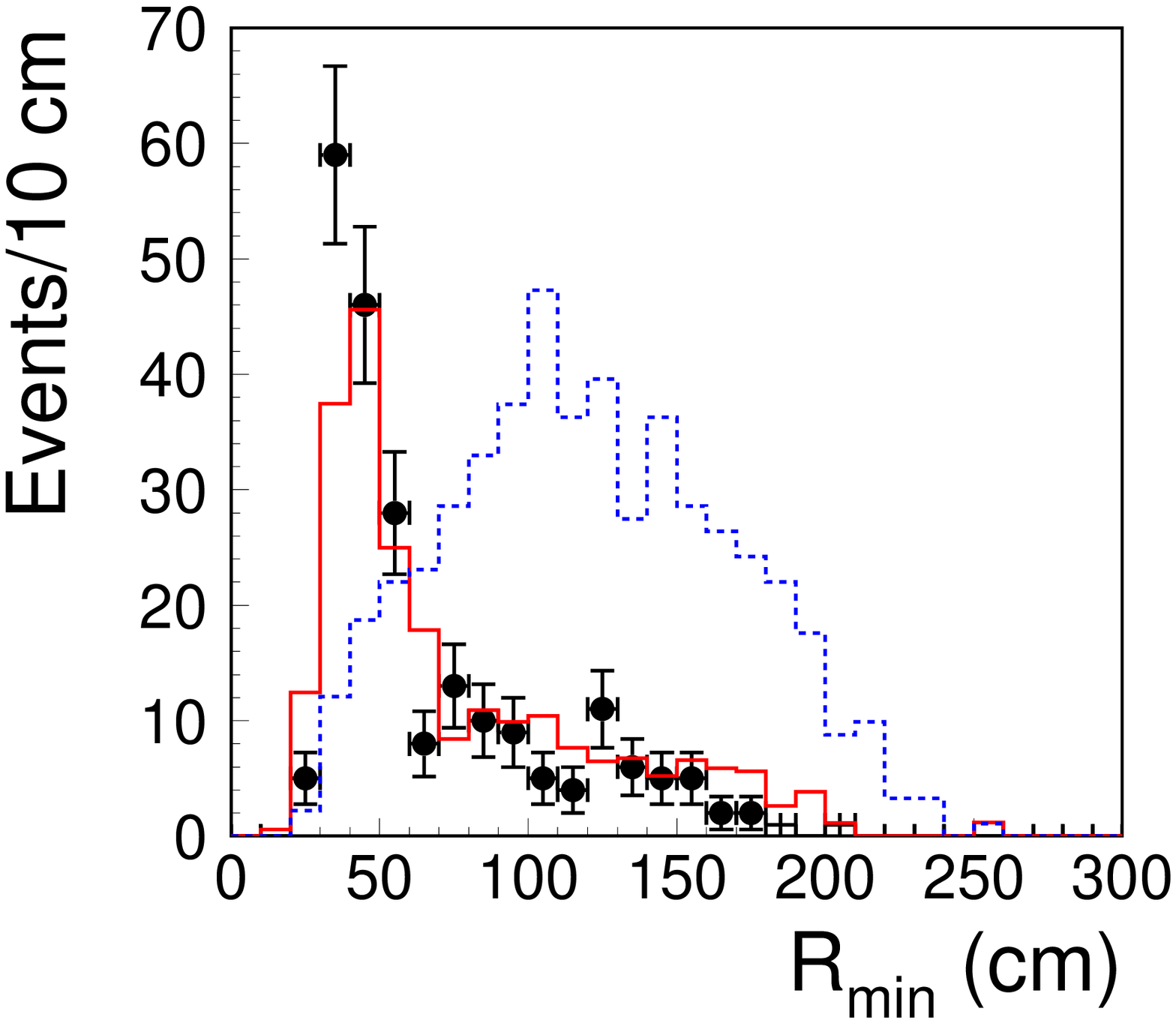}
\includegraphics[width=0.49\textwidth]{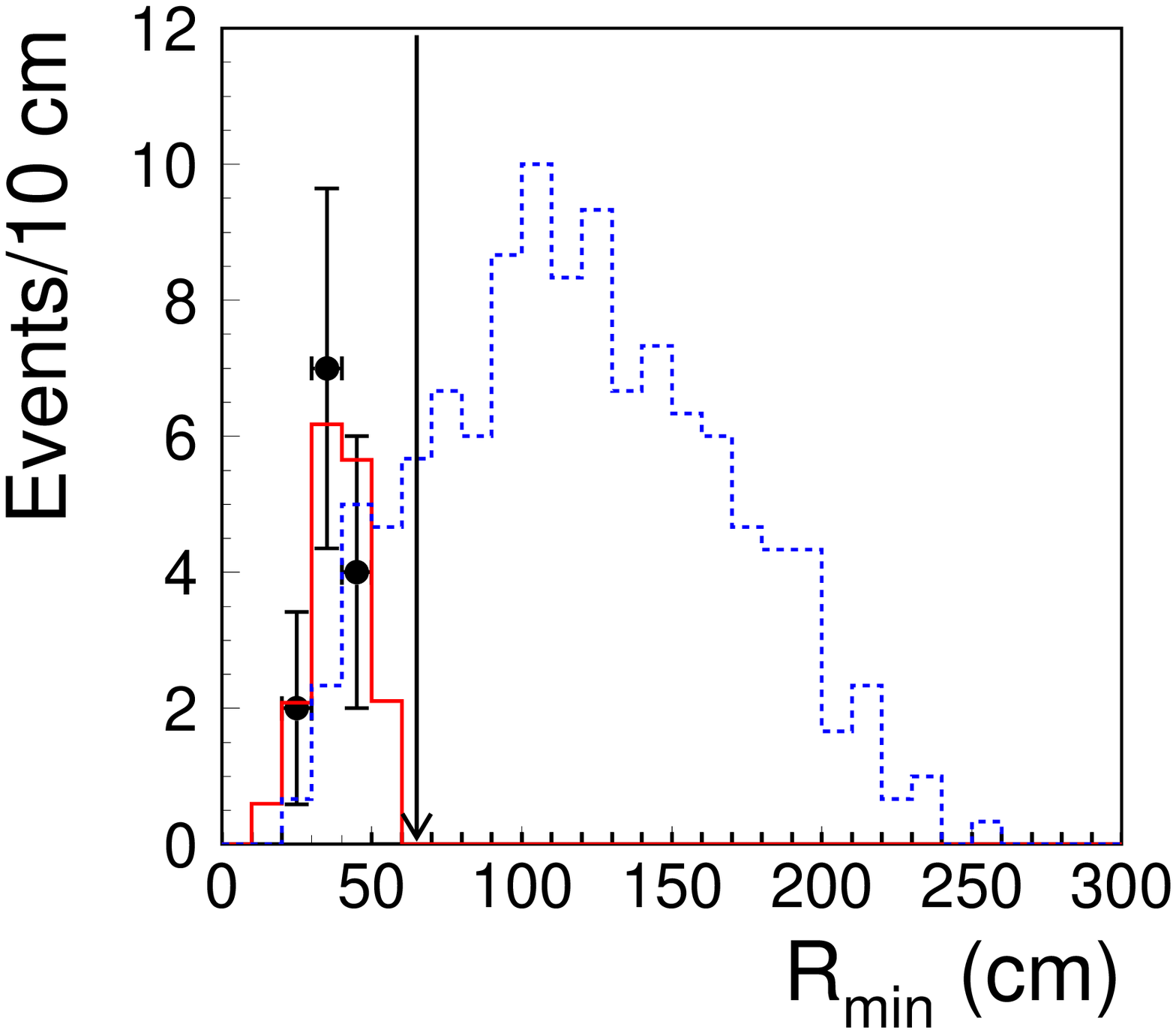}
\end{center}
\caption{Distributions of $R_{min}$ for six-photon events in the signal box applying
the $\chi^2_{fit} <$~57.2 cut (left), and applying $\chi^2_{fit} <$~57.2
and $\Delta > 1.88$ cuts (right).
Black points are data, background simulation is the red histogram.
The dashed histogram represents simulated $K_S \to 3\pi^0$ events.}
\label{fig7}
\end{figure}
At the end of the analysis we find  zero candidates in data and in the
simulated background sample. To assign an error to the Monte Carlo estimate of the background, $N_b$,
we have fit the simulated $R_{min}$ distribution of Fig.~7. (right) with a gaussian and a log-gaussian.
Integrating the events above the cut we estimated $N_b = 0.04 ^{+ 0.15}_{-0.03}$.
\subsection{The normalization sample}
The $K_S \to 2\pi^0$ normalization sample is selected requiring four prompt photons.
The Monte Carlo simulation shows an amount of background of about 0.1$\%$ of the total.
These events are essentially $\phi \to K^+K^-$ decays.
After the $K_L$-crash hard tagging
we find $N_{2\pi} = (7.533 \pm 0.018) \times 10^7$ events.
With the Monte Carlo simulations we have also determined the $K_S \to 2\pi^{0} \to 4\gamma$
efficiency: $\epsilon_{2\pi} = 0.660 \pm 0.002_{stat}$. 
The final number of produced $K_S \to 2\pi^0$ events is:
$ N_{norm} = N_{2\pi}/ \epsilon_{2\pi} = (1.142 \pm 0.005) \times 10^8$.
\subsection{Evaluation of systematic uncertainties}
The systematic uncertainties are related to the number of 
background events and  to the determination of
the acceptance and total efficiencies for the signal,
$\epsilon_{3\pi}$, and normalization, $\epsilon_{2\pi}$,  samples.

For the tagged six-photon sample, we have investigated the uncertainties related to
the observed background at the end of the analysis.
A difference of  $\sim$ 2.4\%  in the EMC energy scale and resolution has been observed between
data and MC simulation and has been  studied using a control sample of $K_S \to 2\pi^0$
events. To evaluate the related systematic uncertainty on the background, we have
repeated the upper limit evaluation with several values of the energy
scale correction in the range of 2.2\%-2.6\%. Similarly, the analysis has been repeated
modifying the resolution used in the definition
of $\zeta_{2\pi}$ and $\zeta_{3\pi}$.
Moreover, we have varied of 1 $\sigma$ the resolution used in the $\Delta$ variable calculation
and removed a data--MC shift correction on $R_{min}$. These variations correspond to a cut change
of 5\% and 6\%, respectively. Similarly, we have removed the data--MC scale correction for $E_{cr}$
and the additional gaussian smearing in the MC $\beta^*$ distribution, both
corresponding to a 5\% variation of the cuts.
The full analysis was repeated in total twenty times applying each
time one of the changes mentioned above. For all of these
checks, we have observed no variation in the number of simulated background.

For the acceptance of  both the signal and normalization
samples, we have evaluated the systematic uncertainty on the photon counting
by comparing data and simulation splitting, accidental probabilities and cluster
reconstruction efficiency.
To determine the probabilities of one, $P_{A1}$, or two, $P_{A2}$,  accidental
clusters in the event we have used out of time clusters originated from earlier bunch
crossing. To estimate the probability of generating one, $P_{S1}$, or more fragments,
$P_{S2}$, per cluster, we have fit the photon multiplicities observed in data using
the experimental values of $P_{A1}$ and $P_{A2}$, and the photon
multiplicities obtained by the simulation~\cite{silarskiPHD,note201}.
Results of these fits are reported in Tab.~\ref{tabprob}.
\begin{table}
\begin{center}
\begin{tabular}{|c|c|c|c|c|}
\hline
 {}& $\boldsymbol{P_{A1}}$ [\%] & $\boldsymbol{P_{A2}}$ [\%] & $\boldsymbol{P_{S1}}$ [\%] & $\boldsymbol{P_{S2}}$ [\%]\\
\hline
\textbf{DATA} & 0.378 $\pm$ 0.004 & 0.025 $\pm$ 0.001 & 0.30 $\pm$ 0.01 & 0.0103 $\pm$ 0.0001\\
\hline
\textbf{MC}& 0.492 $\pm$ 0.004& 0.027 $\pm$ 0.001 & 0.31 $\pm$ 0.01 & 0.0156 $\pm$ 0.0002\\
\hline
\end{tabular}
\end{center}
\caption{
\label{tabprob}
The probabilities to find one ( $P_{A1}$ ) or two ( $P_{A2}$ ) accidental clusters
and to reconstruct one ( $P_{S1}$ ) or more ( $P_{S2}$ ) split clusters
estimated using out of time clusters and fit to the photon multiplicities,
as described in the text.}
\end{table}
The photon reconstruction efficiency, for both data and MC,
was evaluated using a control sample of $\phi \to \pi^+\pi^-\pi^0$ events. The momentum
of one of the photons is estimated from tracking information and position
of the other cluster. The candidate photon is then searched for within a search cone.
The systematic error related to the cluster efficiency has been estimated by
removing the data/MC efficiency correction. The total systematic
uncertainty on  the acceptance for both
measured samples are listed in Tab.~\ref{tabsys}.
Another source of systematic uncertainties originates from the offline filter FILFO~\cite{memo288}
used, during data reconstruction, to reject cosmic rays and machine
background events before starting the track reconstruction. The FILFO efficiency, for both
normalization and signal samples, has been
estimated using the simulation and is very close to 100\%~\cite{silarskiPHD}. 
We have conservatively assigned as systematic uncertainty in data  half of
the difference between the MC evaluated efficiency and 100\%.
We consider completely negligible the influence of trigger efficiency for
both samples, since in~\cite{KLOE_old} it was about 99.5\%
and the $K_L$-crash hard tagging requires a
larger energy release in the calorimeter, which translates in a 
larger trigger efficiency.

The observed difference in the EMC energy scale and resolution between data and simulation
enters also in the $\epsilon_{3\pi}$ evaluation. The effects have been estimated as
$\Delta \epsilon_{3\pi}/\epsilon_{3\pi} = 1.0 $\% from the 	energy scale,
and $\Delta \epsilon_{3\pi}/\epsilon_{3\pi} = 1.1 $\% from the resolution. 
The effect of the cut on $\chi^2_{fit}$ has been tested constructing the ratio between
the cumulative distributions for experimental data and simulation
which leads to a systematics of $\Delta \epsilon_{3\pi}/\epsilon_{3\pi} = 1.46 $\%.
Finally, we have investigated the systematic effect related to
the $R_{min}$ cut by varying its value by 6\%, and estimated its contribution
to be $\Delta \epsilon_{3\pi}/\epsilon_{3\pi} =0.9$\%.
 
All the contributions to the systematic uncertainty are summarized
in Tab.~\ref{tabsys}, with the total systematic uncertainty evaluated adding
all effects in quadrature.
\begin{table}
\begin{center}
\begin{tabular}{|c|c|c|}
\hline
\textbf{Source} & $\boldsymbol{\Delta \epsilon_{2\pi}/\epsilon_{2\pi}~[\%]}$ & $\boldsymbol{\Delta \epsilon_{3\pi}/\epsilon_{3\pi}~[\%]}$\\
\hline
Acceptance & 1.60 & 0.21\\
\hline
Offline filter &0.46 & 0.30\\
\hline
Calorimeter energy scale & -- & 1.00\\
\hline
Calorimeter energy resolution & -- & 1.10\\
\hline
$\chi^2_{fit}$ cut & -- & 1.46\\
\hline
$R_{min}$ cut & -- & 0.90\\
\hline
\bf{TOTAL} & \textbf{1.65} & \textbf{2.30}\\
\hline
\end{tabular}
\end{center}
\caption{
\label{tabsys}
Summary table of the systematic uncertainties on the total
efficiencies for the signal, $\epsilon_{3\pi}$,
and  normalization samples, $\epsilon_{2\pi}$.}
\end{table}
\section{Results}
No events were observed on data in the signal region. 
Equally, no background events are found in the MC simulation based on twice the data statistics.
In the conservative assumption of no background, we estimate an upper
limit on the expected number of signal events UL(Nev$(\ks \to  3
\pio)$) = 2.3  at 90\% ~C.L.,   with a signal efficiency of
$\epsilon_{3\pi} = 0.233 \pm 0.012_{stat} \pm 0.006_{sys}$.
In the same tagged sample we count $N_{norm} = (1.142 \pm 0.005) \times 10^8$
$\ks \to 2 \pio$ events.\\
Systematic uncertainties on background  determination, as well as on
the efficiency evaluation for the signal and normalization samples,
are  negligible in the calculation of the limit.

Using the value BR($\ks \to 2 \pio)= 0.3069 \pm 0.0005$~\cite{pdg2012} we obtain:
\begin{equation}
BR(\ks\to 3 \pio) \leq 2.6 \times 10^{-8}\; \;\;\; {\rm at}\;\; \; 90\%\;\;\; {\rm C.L.}
\end{equation}
which represents the best limit on this decay, improving 
by a factor of $\sim$~5 previous result \cite{KLOE_old}.\\
This result can be translated into a limit on $|\eta_{000}|$: 
\begin{equation}
|\eta_{000}| = \left|\frac{A(K_S \to 3\pi^0)}{A(K_L \to 3\pi^0)}\right| = \sqrt{\frac{\tau_L}{\tau_S}
\frac{BR(K_S \to 3\pi^0)}{BR(K_L \to 3\pi^0)}} \leq 0.0088\;\;\; {\rm at}\;\; \; 90\%\;\;\; {\rm C.L.}
\end{equation}

This describes a circle of radius 0.0088 centered at zero in the $\Re({\eta_{000}})$, $\Im{(\eta_{000})}$
plane and  represents a limit two times smaller than previous result~\cite{KLOE_old}.
\section*{Acknowledgments}
We warmly thank our former KLOE colleagues for the access to the data
collected during 
the KLOE data taking campaign.
We thank the DA$\Phi$NE team for their efforts in maintaining low
background running conditions and their 
collaboration during all data taking. We want to thank our technical staff: 
G.F. Fortugno and F. Sborzacchi for their dedication in ensuring efficient operation of the KLOE computing facilities; 
M. Anelli for his continuous attention to the gas system and detector safety; 
A. Balla, M. Gatta, G. Corradi and G. Papalino for electronics maintenance; 
M. Santoni, G. Paoluzzi and R. Rosellini for general detector support; 
C. Piscitelli for his help during major maintenance periods.\\
We acknowledge the support of the European Community-Research
Infrastructure Integrating Activity `Study of Strongly Interacting Matter'
(acronym HadronPhysics2, Grant Agreement No. 227431) under the Seventh Framework
Programme of EU.
This work was supported also in part by the EU Integrated Infrastructure
Initiative Hadron Physics 
Project under contract number RII3-CT-2004-506078; by the European
Commission under the 7th 
Framework Programme through the `Research Infrastructures' action of
the `Capacities' Programme, Call: 
FP7-INFRASTRUCTURES-2008-1, Grant Agreement No. 283286; by the Polish
National Science Centre through the 
Grants Nos. 0469/B/H03/2009/37, 0309/B/H03/2011/40, 
DEC-2011/03/N/ST2/02641, \\2011/01/D/ST2/00748, 2011/03/N/ST2/02652,
2011/03/N/ST2/02641 and by the Foundation for Polish Science through the MPD programme 
and the project HOMING PLUS BIS/2011-4/3.

\end{document}